\documentstyle[amssymb,eqsecnum,aps,twocolumn]{revtex}

\begin{document}
\draft
\title{Bayesian Methods for Cosmological Parameter Estimation from Cosmic Microwave
Background Measurements}
\author{Nelson Christensen$^1$ and Renate Meyer$^2$ \cite{email}}
\address{\
$^{1}$Physics and Astronomy, Carleton College, Northfield, Minnesota, 55057,
USA \
$^{2}$ Department of Statistics, The University of Auckland, Auckland, New
Zealand\
\
}
\date{\today}
\maketitle

\begin{abstract}
We present a strategy for a statistically rigorous Bayesian approach to the
problem of determining cosmological parameters from the results of
observations of anisotropies in the cosmic microwave radiation background.
We propose the application of Markov chain Monte Carlo methods, specifically
the Metropolis-Hastings algorithm, to estimate the parameters. A complete
statistical analysis is presented, with the Metropolis-Hastings algorithm
described in detail.
\end{abstract}



\narrowtext

\section{Introduction}

Recent observations of the angular power spectrum of the anisotropy of the
cosmic microwave background (CMB) \cite
{boom1,QMAP,QMAP2,QMAP3,MSAM,TOCO1,TOCO2,PythV,Viper} have created much
excitement. This data may be used to estimate cosmological parameters. The
maximum in the angular power spectrum around the multipole value $%
l\thickapprox 200$ is consistent with inflation and a flat universe \cite
{boom1,Knox}. However the apparent absence of \ a peak at $l\thickapprox
400-500$ \cite{boom1} may imply that we have underestimated the size of such
cosmological parameters as the amount of normal baryonic matter or cold dark
matter. This data, coupled with recent supernovae observations \cite{nova},
seems to be leading us to seriously consider the existence of a cosmological
constant. Future observations will provide valuable cosmological information 
\cite{MAP,Planck}.

The extraction of cosmological parameters from the CMB\ anisotropy data is a
complicated and computer intensive activity. This is especially true as the
number of cosmological parameters increases as the modal complexities grow.
Current analysis exercises have included up to ten parameters; the scaler
quadrupole and gravity wave perturbation normalizations ($A_{s}$ and $A_{t}$%
), the scaler and tensor power-law indices for primordial perturbations ($%
n_{s}$ and $n_{t}$), the reionization optical depth $\tau $, the spatial
curvature $\Omega _{k}$, the energy densities for baryonic matter ($\Omega
_{b}$), cold dark matter ($\Omega _{cdm}$), neutrinos ($\Omega _{\nu }$) and
the vacuum ($\Omega _{\Lambda }$) \cite{Max}. Other logical cosmological
parameters could include the Hubble constant, number of neutrino families
and their masses, etc.

Parameter estimation can be comprehensively described within the language of
Bayesian statistics. Application of Bayes' theorem is well suited to
astrophysical observations \cite{Lor}. In Bayesian data analysis the model
consists of a joint distribution over all unobserved (parameters) and
observed (data) quantities. One conditions on the data to obtain the
posterior distribution of the parameters. The starting point of the Bayesian
approach to statistical inference is setting up a full probability model
that consists of the {\em joint}\/ probability distribution of all
observables, denoted by ${\bf z}=(z_{1},\ldots ,z_{n})$ and unobservable
quantities, denoted by $\mbox{\boldmath $\theta$}=(\theta _{1},\ldots
,\theta _{d})$. Using the notion of conditional probability, this joint PDF $%
f({\bf z},\mbox{\boldmath $\theta$})$ can be decomposed into the product of
the PDF of all unobservables, $f(\mbox{\boldmath $\theta$})$, referred to as
the {\em prior}\/ PDF of $\mbox{\boldmath $\theta$}$, and the conditional
PDF of the observables given the unobservables, $f({\bf z}|%
\mbox{\boldmath
$\theta$})$, referred to as the sampling distribution or {\em likelihood},
i.e.\ 
\[
f({\bf z},\mbox{\boldmath $\theta$})=f(\mbox{\boldmath $\theta$})f({\bf z}|%
\mbox{\boldmath $\theta$}). 
\]
The prior PDF contains all the information about the unobservables that is
known from substantive knowledge and expert opinion {\em before} observing
the data. All the information about the $\mbox{\boldmath $\theta$}$ that
stems from the experiment is contained in the likelihood. In the light of
the data, the Bayesian paradigm then updates the prior knowledge about $%
\mbox{\boldmath $\theta$}$, $f(\mbox{\boldmath $\theta$})$, to the {\em %
posterior} PDF of $\mbox{\boldmath $\theta$}$, $f(\mbox{\boldmath $\theta$}|%
{\bf z})$. This is done via an application of Bayes theorem through
conditioning on the observations 
\[
f(\mbox{\boldmath $\theta$}|{\bf z})=\frac{f(\mbox{\boldmath $\theta$},{\bf z%
})}{m({\bf z})}\propto f(\mbox{\boldmath $\theta$})f({\bf z}|%
\mbox{\boldmath
$\theta$}) 
\]
where $m({\bf z})=\int f({\bf z}|\mbox{\boldmath $\theta$})f(%
\mbox{\boldmath
$\theta$})d\mbox{\boldmath $\theta$}$ is the marginal PDF of ${\bf z}$ which
can be regarded as a normalizing constant as it is independent of $%
\mbox{\boldmath $\theta$}$. The Bayesian approach is based on the likelihood
function but treats the parameters as random variables and assumes a joint
prior distribution that summarizes the available information about the
parameters before observing the data. In the light of the observations, the
information about the unknown parameters is then updated via Bayes theorem
to the posterior distribution which is proportional to the product of
likelihood and prior density \cite{carl96}.

The problem of marginalization of parameters has been the Achilles heel of
Bayesian approaches to parameter estimation. The inevitable difficulty in
calculating the multidimensional integrals necessary for the determination
of posterior probability distributions has hindered efforts. However, these
impediments have been overcome by the progress made within the last decade
in Bayesian computational technology via Markov chain Monte Carlo (MCMC)
methods \cite{gilk96}. Since its initial application in digital signal
analysis \cite{Gem} MCMC\ methods have revolutionized many areas of applied
statistics and should have an impact for cosmological parameter estimation
from CMB measurements.

In this paper we do NOT implement a MCMC method for estimating cosmological
parameters. This task is best left to the active scientists in the field.
Instead we describe how researchers may adapt their existing specialized
software in order to produce probability distributions functions for the
parameters that have been derived in a statistically rigorous manner.

In section II we review methods that have been used to estimate cosmological
parameters from CMB\ measurements, plus techniques for calculating
likelihoods. In section III we describe the Bayesian approach to statistical
inference and its implementation via Markov chain Monte Carlo methods. In
section IV we describe a methods for applying MCMC\ methods to cosmological
parameter estimation with CMB\ data. Section V presents our conclusions.

\section{Current Statistical Methods with CMB\ Data}

\subsection{Parameter Estimation}

Various statistical approaches have attempted to estimate cosmological
parameters from the CMB\ data. The maximum likelihood is an example \cite
{Maxima1,boom2,Knox2}. Likelihood techniques aim at finding the values of
the parameters that maximize the likelihood function. However, the maximum
of the likelihood is not necessarily the peak value of the parameter's
marginal distribution (its PDF) or its mean. The maximum likelihood
parameter estimates will differ from the Bayes estimators. The maximum
likelihood is not always good for parameter estimation; for a given
parameter, the maximization depends on whether or not other parameters have
been integrated out. A proper application of Bayes' theorem is attempted in
some studies through the inclusion of prior distributions \cite
{boom1,Max,boom2,boom3}. The typical approach is to select a range for the
parameters and limit the maximum likelihood technique to that region in
parameter space. Sophisticated {\it frequentist} techniques are implemented
in order to maximize the likelihood, such as the {\it Levenberg-Marquardt
method} \cite{BJK98,NumR}, or the {\it downhill simplex method} \cite
{boom2,NumR}.

In the methods described above the problem of maginalizing over parameters
is non-trivial. Typically the approach is to marginalize over all parameters
except the vacuum energy density ($\Omega _{\Lambda }$) and the matter
energy density ($\Omega _{m}=\Omega _{b}+\Omega _{cdm}$). This procedure
gets prohibitively difficult as the number of parameters increases; in some
cases the amount of computation can be expected to rise exponentially with
parameter number. Hence, the techniques that have already been used will be
difficult to apply to cosmological models with increasing complexity.
Marginalizing over parameters will also become difficult if one no longer
assumes that the variables are log-normally distributed, or if the
cosmological CMB\ anisotropy signal is actually non-Gaussian \cite
{non-Gauss,Zald}.

Numerical maximization techniques used in maximum likelihood estimation,
such as the Levenberg-Marquardt method, are only guaranteed to find a local
maximum. Once they reached a local maximum they might get stuck in their
search and not reach the global maximum. For this reason statisticians have
applied {\em simulating annealing}, which is a technique for global
optimization. Simulated annealing has been attempted for cosmological
parameter estimation \cite{Hann,Knox3}. Simulated annealing is related to
MCMC as its core component is the Metropolis Hastings algorithm. In this
method the parameter space is searched in a random way. A new parameter
space point is reached with a probability that depends on the likelihood and
an {\it effective temperature} term. In the limit where the temperature
approaches zero the {\it thermodynamics} of this parameter space search
finds the system approaching the maximum of the likelihood. Stochastic
methods such as these substitute deterministic integration by a statistical
estimation problem. Although these methods are applicable to
high-dimensional problems, they can be very inefficient in certain
situations. The shortcomings of simulating annealing are that there is no
guarantee that it will actually find the global maximum in finite time. The
efficiency depends very much on specifying a good cooling schedule which
involves the arbitrary and skillful choice of various cooling parameters
that specify the algorithm and determine its performance.

\subsection{Calculating the Likelihood}

Implicit in the problem of parameter estimation is the calculation of the
likelihood, namely the probability of obtaining the data given a particular
model (and its associated parameters). The sizes of the data sets from CMB\
experiments are immense, so it is impractical to use all the raw data for
producing the likelihood. Instead, radical compression of the data is
imperative \cite{BJK98}. Even so, a precise derivation of the likelihood is
very difficult \cite{Max,BJK98,Bart1,BJK00}. Approximate techniques have
been developed that take as input the confidence intervals of the
experimental results \cite{BJK98,Bart1,BJK00,Bart2,MADCAP}. Software for
calculating the likelihoods, RADPACK is freely available at {\it %
http://flight.uchicago.edu/knox/radpack.html} \cite{MADCAP}.

In order to calculate the likelihood it is necessary to have the angular
power spectrum of this CMB\ anisotropy for a specific model (with its
associated cosmological parameters). This calculational task is accomplished
by code such as CMBfast \cite{CMBFAST} or CAMB \cite{CAMB}. This code
accepts the cosmological parameters as input, and returns the angular power
spectrum of the CMB\ anisotropies, $C_{l}$. These software packages serve as
the work-horses of current CMB\ statistical work. For example, the
likelihood has been calculated 30,311,820 times in order to cover a region
in a ten-dimensional cosmological parameter space \cite{Max}. In other
studies the likelihood was evaluated as needed within the calculation \cite
{Knox2,boom2}.

With a large number of parameters it becomes impractical to marginalize by
integration. So when attempts are made to produce constraints in the $\Omega
_{\Lambda }-\Omega _{m}$ plane approximate marginalization techniques must
be applied. One can estimate the parameters by accepting the values from the
maximum likelihood on the parameter grid \cite{Line,Max2}, and then
interpolate between grid points with a{\em \ cubic spline} \cite{Max}.
Another technique is to maximize the likelihood with respect to all
remaining parameters for a fixed point in the $\Omega _{\Lambda }-\Omega
_{m} $ plane \cite{Knox2,boom2}. These methods are further complicated due
to degeneracies among the cosmological parameters.\cite{Efst}

\section{Bayesian Posterior Computation via MCMC}

The difficulty with the Bayesian approach to parameter estimation is
high-dimensional integration. To calculate the normalizing constant of the
joint posterior PDF, for instance, requires $d$-dimensional integration.
Having obtained the joint posterior PDF of $\mbox{\boldmath $\theta$}$, the
posterior PDF of a single parameter $\theta _{i}$ of interest can be
obtained by integrating out all the other components, i.e. 
\[
f(\theta _{i}|{\bf z})=\int \ldots \int f(\mbox{\boldmath $\theta$}|{\bf z}%
)d\theta _{1}\ldots d\theta _{i-1}d\theta _{i+1}\ldots d\theta _{d}. 
\]
Calculation of the posterior mean of $\theta _{i}$ necessitates a further
integration, e.g.\ $E[\theta _{i}|{\bf z}]=\int \theta _{i}f(\theta _{i}|%
{\bf z})d\theta _{i}$.

As the joint posterior is too complex to sample from directly, we propose to
use a MCMC method \cite{gilk96,chri98}. Instead of generating a sequence of
independent samples from the joint posterior, in MCMC a Markov chain is
constructed, whose equilibrium distribution is just the joint posterior.
Thus, after running the Markov chain for a certain ''burn-in'' period, one
obtains (correlated) samples from the limiting distribution, provided that
the Markov chain has reached convergence.

One method for generating a Markov chain is via the Metropolis-Hastings
algorithm. The MH algorithm was developed by Metropolis et al.\ \cite{metr53}
and generalized by Hastings \cite{Hast}. It is a MCMC method which means
that it generates a Markov chain whose equilibrium distribution is just the
target density, here the joint posterior PDF $f(\mbox{\boldmath $\theta$}|%
{\bf z})$. For ease of notation, let us assume that the target density is
some $f(x)$, where $x$ can be $d$-dimensional. Then a Markov chain is
specified by constructing a transition kernel $P(x,A)$, giving the
conditional probability to move from state $x$ to a point in $A\in {\cal B}$%
, the Borel $\sigma $-field on $I\!\!R^{d}$. As the probability to stay in
the current state $x$, $P(x,\{x\})$, is not necessarily 0, we assume that
the transition kernel can be expressed as 
\[
P(x,dy)=p(x,y)dy+r(x)\delta _{x}(dy)
\]
where $p(x,x)=0$ and $\delta _{x}(dy)=1$ if $x\in dy$, 0 otherwise, and $%
r(x)=1-\int p(x,y)dy$. It is well known that if $p(x,y)$ satisfies `{\em %
reversibility}' or `{\em detailed balance}', i.e.\ 
\[
f(x)p(x,y)=f(y)p(y,x)
\]
then $f(x)$ is the invariant density of the Markov chain. Together with {\em %
irreducibility} and {\em aperiodicity}, this gives a sufficient condition
for the convergence of the Markov chain to its stationary distribution, see
Tierney \cite{Luke} for further details.

The MH algorithm shares the concept of a generating density with the
well-known simulation technique of {\em rejection sampling}. However, the 
{\em candidate generating density} $q(y|x)$, $\int q(y|x)dy=1$, can now
depend on the current state $x$ of the sampling process, and instead of
rigorously accepting or rejecting a new candidate $y$, it is accepted with a
certain {\em acceptance probability} $\alpha (y|x)$ also depending on the
current state $x$, and chosen such that the transition probability $%
p(x,y)=q(y|x)\alpha(y|x)$ satisfies detailed balance. This is met by setting 
\[
\alpha (y|x))=\min \left\{ \frac{ f(y) q(x|y)}{f(x) q(y|x) },1\right\} 
\]
if $f(x)q(y|x))>0$ and $\alpha(y|x)=1$ otherwise. Note that irreducibility
is guaranteed if $q(y|x)$ is positive on the support of $f$.\bigskip

The steps of the MH algorithm are therefore:\bigskip

\begin{tabular}{ll}
Step 0: & Start with an arbitrary value $x_0$ \\ 
&  \\ 
Step $i+1$: & Generate $y$ from $q(.|x_i)$ and $u$ from $U(0,1)$ \\ 
& If $u\leq \alpha (y|x_i)$ set $x_{i+1}=y$ (acceptance) \\ 
& If $u>\alpha (y|x_i)$ set $x_{i+1}=x_{i}$ (rejection)
\end{tabular}
\bigskip

The MH algorithm does not require the normalization constant of the target
density. The outcomes from the MH algorithm can be regarded as a sample from
the invariant density only after a certain `burn-in' period. For issues
concerning convergence diagnostics, the reader is referred to Cowles and
Carlin \cite{CC}. Note that an important special case of the MH algorithm is
the `{\em independence chain}' where $q(y|x)=q(y)$, i.e.\ a new candidate is
generated independently of the current state $x$.

Various methods to assess convergence, i.e.\ methods used for establishing
whether an MCMC algorithm has converged and whether its output can be
regarded as samples from the target distribution of the Markov chain, have
been developed and implemented in CODA \cite{coda95}. CODA is a menu-driven
collection of SPLUS functions for analyzing the output of the Markov chain.
Besides trace plots and the usual tests for convergence, CODA calculates
statistical summaries of the posterior distributions and kernel density
estimates.

MCMC techniques have been applied in numerous areas, from science to
economics. Applications of state-space modeling in finance, e.g. stochastic
volatility models applied to time series of daily exchange rates or returns
of stock exchange indices, easily have 1000-5000 parameters and the Gibbs
sampler shows slow convergence due to high posterior correlations \cite
{meye00,jacq94,meyu00}. Specially tailored MCMC algorithms, like multi-move
Gibbs samplers or Metropolis-Hastings algorithms, can markedly improve the
speed of convergence \cite{kim98}.

\section{Applying MCMC\ Methods to CMB\ Anisotropy Data}

It is possible for researchers to generate probability density functions for
cosmological parameters from the CMB\ anisotropy data. This can be done in a
Bayesian fashion with the MCMC serving as a means of conducting a proper
marginalization over parameters.\ The implementation of the MCMC method
would be relatively straight forward. Instead of calculating the likelihood
at uniform locations in the parameter space \cite{Max}, one would let the
MCMC do its intelligent walk through the space. Uniform {\em a priori}
distributions for the parameters seems reasonable, so the MCMC\ would sample
the parameter space defined by them. Since the likelihood function can not
be written explicitly in terms of the cosmological parameters, but instead
in terms of the CMB\ anisotropy power spectra terms $C_{l}$, it will be
necessary to implement a Metropolis-Hastings MCMC routine.

The Markov chain would commence at a randomly selected position in parameter
space $(\theta _{1}^{(0)},\ldots ,\theta _{d}^{(0)})$. With the parameter
set one would then utilize CMBfast \cite{CMBFAST} or CAMB \cite{CAMB} to
generate a set of CMB anisotropy angular power spectrum components ($C_{l}$%
),.and a likelihood would then be calculated with CMB\ anisotropy data \cite
{BJK98,Bart1,Bart2,MADCAP}. New values for the parameters $(\theta
_{1}^{(1)},\ldots ,\theta _{d}^{(1)})$ would be selected via sampling from
the a priori distributions. However, these values would not necessarily be 
{\em accepted} as new values. First they would be used as input to CMBfast 
\cite{CMBFAST} or CAMB \cite{CAMB} to generate a set of CMB anisotropy
angular power spectrum components ($C_{l}$), and a likelihood would then be
calculated with CMB\ anisotropy data \cite{BJK98,Bart1,Bart2,MADCAP}. The
new values would be {\em accepted} or rejected according the to following
test; a random number, $u$, would be generated between 0 and 1. If \ $u\leq
\min \left[ 1,f({\bf z}|\theta _{1}^{(1)},\ldots ,\theta _{d}^{(1)})/f({\bf z%
}|\theta _{1}^{(0)},\ldots ,\theta _{d}^{(0)})\right] $ (where $f({\bf z}%
|\theta _{1}^{(1)},\ldots ,\theta _{d}^{(1)})$ is the likelihood in terms of
data ${\bf z}$ and cosmological parameters $(\theta _{1},\ldots ,\theta _{d})
$) then the new parameters is {\em accepted} into the chain, if not the next
chain element has values equal to that of the previous state. A new set of
parameters would then be randomly sampled from the a priori distributions
and the procedure would continue.

The generated chain of parameter values would form the set from which the
statistical properties would be derived. After running the Markov chain for
a certain ''burn-in'' period (in order for the Markov chain to reach
convergence) one obtains (correlated) samples from the limiting
distribution. This process would continue for a sufficiently long time (as
determined by convergence diagnostics \cite{coda95}).

After the {\em burn-in} the frequency of appearance of parameters would
represent the actual posterior density of the parameter. From the posterior
density one can then create confidence intervals. Summary statistics are
produced from the distribution, such as posterior mean and standard
deviation. A cross-correlation matrix is also easily produced; this would be
of great importance for quantifying the ''degeneracy'' of $\Omega _{m}$ and $%
\Omega _{m}$ \cite{Efst}. A distinct advantage of the MCMC\ approach is that
computational time scales linearly with parameter number. Hence, the MCMC
approach to cosmological parameter estimation may provide the best strategy
when testing complex models with numerous parameters. Also, MCMC\ methods
could work with the exact form of the likelihood, although the approximate
forms may prove to be sufficiently adequate \cite{BJK98,Bart1,Bart2,MADCAP}.

The above method would constitute the simplest implementation of the
Metropolis-Hastings method, that of an independence chain. We have just used
an {\em acceptance probability} $\alpha (\mbox{\boldmath $\theta$}^*|%
\mbox{\boldmath $\theta$})$, defined by 
\[
\alpha (\mbox{\boldmath $\theta$}^*|\mbox{\boldmath $\theta$})=\min \left\{ 
\frac{f({\bf z}|\mbox{\boldmath $\theta$}^*)q(\mbox{\boldmath $\theta$}|%
\mbox{\boldmath $\theta$}^*)}{f({\bf z}|\mbox{\boldmath $\theta$})q(%
\mbox{\boldmath $\theta$}^*|\mbox{\boldmath $\theta$})},1\right\} 
\]
where the generating density $q(\mbox{\boldmath $\theta$}^*|%
\mbox{\boldmath
$\theta$})$ is the uniform density over the parameter space and thus, in
particular, independent of the current state. By using uniform priors, the
posterior PDFs in the acceptance probability calculation reduce to the
likelihoods. However, the efficiency of a Metropolis-Hastings algorithm
depends crucially on the form of the generating density $q(%
\mbox{\boldmath
$\theta$}^*|\mbox{\boldmath $\theta$})$. Just using a uniform distribution
that does not even depend on the current state $\theta $ is the most simple
but probably most inefficient way to accomplish the task. Even with a
uniform distribution, the algorithm will be irreducible/aperiodic/reversible
and thus the Markov chain will converge towards its stationary distribution.

A slightly better way might be to use a uniform distribution in a
neighborhood of the current $\mbox{\boldmath $\theta$}$. Any prior
information could be useful, such as correlations that one could use to
specify a multivariate normal centered around the current $%
\mbox{\boldmath
$\theta$}$ with a covariance matrix that takes said correlations into
account. The optimization of the Metropolis-Hastings MCMC strategy will
inevitably require experimentation with the generating density $q(%
\mbox{\boldmath $\theta$}^{\ast }|\mbox{\boldmath $\theta$})$. While this
may require some detailed study, the benefit will be the ability to generate
posterior distributions for a large number of cosmological parameters.

\section{Discussion}

A\ proper Bayesian approach to parameter estimation allows one to
simultaneously estimate both dynamic signals and measurement noise. MCMC
methods have demonstrated their importance in Bayesian parameter estimation
problems with large numbers of parameters. As cosmological models grow in
complexity it will become necessary to use techniques such as those
discussed here in order to handle marginalization of parameters.
Furthermore, a MCMC\ approach is not restricted to the assumption of
Gaussian noise. A heavy-tailed observation error distribution such as a
Student-$t$-distribution with large degrees of freedom might be more
appropriate to allow for crude measurement errors and ensure that resulting
estimates will be robust against additive outliers. Non-Gaussian error
distributions are readily incorporated into MCMC calculations. We therefore
strongly advocate the Bayesian approach via Metropolis-Hastings sampling in
future analyses of cosmological parameter estimation from CMB\ data.

\bigskip

\acknowledgments
This work was supported by the Royal Society of New Zealand Marsden Fund,
the University of Auckland Research Committee, and Carleton College.

\appendix

\end{document}